\documentclass[aps,preprint]{revtex4}%
\usepackage{amsfonts}
\usepackage{amsmath}
\usepackage{amssymb}
\usepackage{graphicx}%
\providecommand{\U}[1]{\protect\rule{.1in}{.1in}}
\begin{document}
\title[ ]{Helium Ground State Treated in Classical Physics with Classical Zero-Point Radiation}
\author{Timothy H. Boyer}
\affiliation{Department of Physics, City College of the City University of New York, New
York, New York 10031}
\keywords{}
\pacs{}

\begin{abstract}
The ground State of the helium atom is considered within classical
electrodynamics which includes classical electromagnetic zero-point radiation.
\ Approximate energy balance between energy loss through emitted radiation and
energy gain from classical zero-point radiation is found when the two
electrons are treated as charges located on opposite ends of a diameter of a
common circular orbit around the nucleus. \ The classical result gives
approximately the same value as that given by the quantum calculation, but is
a different value. \ 

\end{abstract}
\maketitle

\section{Introduction}

Old quantum theory seemed to work very well for relativistic hydrogen, but it
seemed unable to account for the ground state of helium.\cite{Pais215}
\ Indeed, this failure was one of the reasons why old quantum theory was
abandoned. \ Thus when modern quantum mechanics with its absence of classical
trajectories was suggested, it was welcome by those working on ideas of atomic
physics. \ Old quantum theory indeed assigned integer values to the action
variables, but it did not envision resonant behavior between particle motion
and any source of random Brownian-like behavior. Here we point out a classical
solution to the old helium problem. \ Classical electrodynamics, in which
classical zero-point radiation is present, will indeed account for the
ionization energy of ground-state helium. \ 

\section{Basic Ideas}

\subsection{Helium}

Helium consists of a heavy nucleus with two positive electronic charges, +2e,
outside of which are two electrons, each of which has one negative electronic
charge $-e$. \ The most symmetrical arrangement corresponds to the two
electrons located on opposite ends of a diameter of a circular orbit with the
nucleus at the center of the orbit. \ Since the electrons repel each other,
they would tend to stay on opposite ends of an orbital diameter. \ If we take
the nucleus as the origin of our coordinates, this arrangement of charges
gives the neutral atom an electric  quadrupole moment, but no electric dipole
moment. \ 

In the case of hydrogen,\cite{B} the classical analysis involved the electric
dipole moment, and, less importantly, higher multipole moments. \ For helium,
the radiation contributions from the dipole terms for the two charges cancel
each other. \ Therefore for helium, we must start with the quadrupole moment
for both the emission of radiation energy and the gain of radiation energy
from zero-point radiation. \ 

\section{Classical Electromagnetic Radiation}

One of the puzzling features of Bohr's assumptions\cite{Bohr} in 1913 was the
stipulation that the electron moving in a circular obit in the hydrogen atom
did not radiate. \ Such an assumption is contrary to classical
electrodynamics. \ Here do not make Bohr's assumption, but rather keep the
classical idea of a charged particle losing energy due to emission of
radiation. \ However, average energy balance is maintained due to a gain of
orbital energy from random classical electromagnetic zero-point radiation. \ 

The source-free random classical electromagnetic radiation in a very large
\textit{spherical} cavity of radius $\mathsf{R}$ can be written
as\cite{Jackson}%

\begin{align}
\mathbf{E}(\mathbf{r,}t)  &  =\operatorname{Re}\sum\nolimits_{n=1}^{\infty
}\sum\nolimits_{l=1}^{\infty}\sum\nolimits_{m=-l}^{m=l}\left\{  \exp\left[
i\left(  -k_{nl}^{M}ct+\theta_{nlm}^{M}\right)  \right]  \left[  ia_{nlm}%
^{M}j_{l}\left(  k_{nlm}^{M}r\right)  \mathbf{X}_{l,m}\left(  \theta
,\phi\right)  \right]  \right. \nonumber\\
&  \left.  +\exp\left[  i\left(  -k_{nlm}^{E}ct+\theta_{nlm}^{E}\right)
\right]  \left[  a_{nlm}^{E}/\left(  -ik_{nlm}^{E}\right)  \right]
\nabla\times\left[  j_{l}\left(  k_{nlm}^{E}r\right)  \mathbf{X}_{lm}\left(
\theta,\phi\right)  \right]  \right\}  , \label{Ezprt}%
\end{align}
and%

\begin{align}
\mathbf{B}(\mathbf{r,}t)  &  =\operatorname{Re}\sum\nolimits_{n=1}^{\infty
}\sum\nolimits_{l=1}^{\infty}\sum\nolimits_{m=-l}^{m=l}\left\{  \exp\left[
i\left(  -k_{nlm}^{E}ct+\theta_{nlm}^{E}\right)  \right]  \left[  ia_{nlm}%
^{E}j_{l}\left(  k_{nlm}^{E}r\right)  \mathbf{X}_{l,m}\left(  \theta
,\phi\right)  \right]  \right. \nonumber\\
&  \left.  +\exp\left[  i\left(  -k_{nlm}^{M}ct+\theta_{nlm}^{M}\right)
\right]  \left[  a_{nlm}^{M}/\left(  ik_{nlm}^{M}\right)  \right]
\nabla\times\left[  j_{l}\left(  k_{nlm}^{M}r\right)  \mathbf{X}_{lm}\left(
\theta,\phi\right)  \right]  \right\}  , \label{Bzprt}%
\end{align}
where $a_{nlm}^{E}$ and $a_{nlm}^{M}$ are the amplitudes of the electric and
magnetic radiation modes, $j_{l}$ is the spherical Bessel function of order
$l$, $\mathbf{X}_{l,m}\left(  \theta,\phi\right)  $ is the vector spherical
harmonic, and $\theta_{nlm}^{E},\theta_{nlm}^{E}$\ are the random phases
distributed uniformly on $(0,2\pi]$ and independently for each radiation
mode.\cite{EH}\cite{Rice} \ The representation of the rotation group refers to
the integers $l$ and $m$ which are unrelated to the frequency $\omega$ of the
normal mode.

Here we have the vector spherical harmonic%
\begin{equation}
\mathbf{X}_{lm}\left(  \theta,\phi\right)  =\frac{\mathbf{L}Y_{lm}\left(
\theta,\phi\right)  }{\sqrt{l\left(  l+1\right)  }}=\frac{\mathbf{r\times
\nabla}Y_{l,m}\left(  \theta,\phi\right)  }{i\sqrt{l\left(  l+1\right)  }},
\end{equation}
where the spherical harmonics are given by%
\begin{equation}
Y_{lm}\left(  \theta,\phi\right)  =\left[  \sqrt{\frac{\left(  2l+1\right)
}{4\pi}\frac{\left(  l-m\right)  !}{\left(  l+m\right)  !}}P_{l}^{m}\left(
\cos\theta\right)  \right]  \exp\left[  im\phi\right]  .
\end{equation}
The only complex part is the exponential behavior in $\phi$, $\exp\left[
im\phi\right]  $. \ 

For random radiation, we have the scale given by\cite{Disguised}%

\begin{equation}
\left\vert a_{nlm}^{E}\right\vert ^{2}=\frac{16\pi\left(  k_{nl}^{E}\right)
^{2}}{\mathsf{R}}\left[  U_{nlm}\left(  k_{nlm}^{E}\right)  \right]  .
\label{aEnlm2}%
\end{equation}

The situation for \textit{magnetic} modes $a_{nlm}^{M}$ is exactly analogous.
\ Magnetic radiation modes\cite{Burko} contribute for a charged particle in a
circular orbit beginning at $l=2$ with a magnitude comparable to the
\textit{electric} multipole of order $l=3$.

For the limit of large radius $\mathsf{R}$ of the enclosing sphere containing
standing waves, we have\cite{Disguised} $dk=\pi dn/\mathsf{R}$ and \
\begin{equation}
\sum\nolimits_{n=1}^{\infty}\rightarrow\int_{0}^{\infty}dn=\int_{0}^{\infty
}dk\frac{\mathsf{R}}{\pi}=\int_{0}^{\infty}d\omega\frac{\mathsf{R}}{\pi
c}.\,\label{SumC}%
\end{equation}
If we use random phases for closely-space radiation modes labeled by $nlm$,
then the amplitude involves the average energy $U_{nlm}.$ \ The number of
normal modes per unit (angular) frequency per unit volume is $\omega
^{2}/\left(  \pi^{2}c^{3}\right)  ,$ which gives an energy per unit angular
frequency interval per unit volume $\left[  \omega^{2}/\left(  \pi^{2}%
c^{3}\right)  \right]  U\left(  \omega\right)  .$ \ 

\subsection{Classical Zero-Point Radiation}

A single plane electromagnetic wave is universally accepted as part of
classical electromagnetism. \ So too are any finite number of plane waves of
differing energy, phase, and direction. \ A Lorentz-invariant spectrum of
plane waves corresponds to an energy per normal mode given by\cite{Marshall}
$U_{nlm}\left(  \omega\right)  =const\times\omega,$ where $const$ can be any
positive numerical constant. \ If we regard the spectrum of random classical
zero-point radiation as the cause of the Casimir forces\cite{Casimir} between
two parallel plates, then the constant must have the numerical value
$const=\left(  1/2\right)  \hbar$ where $\hbar$ is Planck's constant
\begin{equation}
U_{nlm}^{zp}\left(  \omega\right)  =\frac{1}{2}\hbar\omega.
\end{equation}
The spectrum of random Lorentz-invariant classical radiation will account for
both the magnitude and the space-dependent functional form of the force
between the parallel plates.\cite{Cexp}

\section{Gain of Energy for a Charge in a Circular Orbit in Zero-Point
Radiation}

\subsection{Equation of Motion for the Charge Driven by Random Radiation}

The electric dipole field reverses the $\phi$-component on opposite sides of a
circular orbit with the coordinate origin at the center. \ Two charges $e$ at
opposite ends of a diameter for a \textit{circular} orbit in the $xy$-plane
will each experience a force due to the $\phi$-component of the
\textit{quadrupole} electric field $E_{\phi}^{E}(r_{e},\pi/2,\phi_{e}\left(
t\right)  ,t)$ at the angular location $\phi_{e}\left(  t\right)  $ of the
charge. \ For each charge, Newton's second law gives%
\begin{equation}
Mr_{e}^{2}\frac{d^{2}\phi_{e}\left(  t\right)  }{dt^{2}}=r_{e}eE_{\phi}%
^{E}(\mathbf{r}_{e}\left(  t\right)  ,t),\label{Mr2dphi}%
\end{equation}
where $M$ is the nonrelativistic particle mass. \ For the solution of this
equation, we use variation of parameters analogous to Born's work\cite{Born}
for the linear harmonic oscillator. \ Now the \textit{source-free} solutions
of the homogeneous equation $Mr_{0}^{2}\ddot{\phi}=0$ for the circular orbit
are $\phi\left(  t\right)  =a$ and $\phi\left(  t\right)  =bt$ where $a$ and
$b$ are constants. \ Then for each charge, it follows that the particular
solution $\phi_{p}(t)$ of Eq. (\ref{Mr2dphi}) is obtained by variation of
parameters as%

\begin{align}
Mr_{e}\phi_{p}(t) &  =-\int_{0}^{t^{\prime}=t}dt^{\prime}\left[  eE_{\phi}%
^{E}(\mathbf{r}_{e}\left(  t^{\prime}\right)  ,t^{\prime})\right]  t^{\prime
}+\int_{0}^{t^{\prime}=t}dt^{\prime}\left[  eE_{\phi}^{E}(\mathbf{r}%
_{e}\left(  t^{\prime}\right)  ,t^{\prime})\right]  t\nonumber\\
&  =\int_{0}^{t^{\prime}=t}dt^{\prime}\left[  eE_{\phi}^{E}(\mathbf{r}%
_{e}\left(  t^{\prime}\right)  ,t^{\prime})\right]  \left(  t-t^{\prime
}\right)  .\label{Mrphip}%
\end{align}
Note that the angular velocity $d\phi_{p}\left(  t\right)  /dt$ follows as%
\begin{align}
Mr_{e}\frac{d\phi_{p}(t)}{dt} &  =-\left[  eE_{\phi}^{E}(\mathbf{r}_{e}\left(
t\right)  ,t)\right]  t+\left[  eE_{\phi}^{E}(r_{e},\pi/2,\phi_{e}\left(
t\right)  ,t)\right]  t\nonumber\\
&  +\int_{0}^{t^{\prime}=t}dt^{\prime}\left[  eE_{\phi}^{E}(\mathbf{r}%
_{e}\left(  t^{\prime}\right)  ,t^{\prime})\right]  \nonumber\\
&  =\int_{0}^{t^{\prime}=t}dt^{\prime}\left[  eE_{\phi}^{E}(\mathbf{r}%
_{e}\left(  t^{\prime}\right)  ,t^{\prime})\right]  .\label{Mr0v}%
\end{align}
And indeed the equation of motion (\ref{Mr2dphi}) is satisfied, since when we
take the second derivative with respect to time $t$, we obtain \
\begin{equation}
Mr_{e}\frac{d^{2}\phi_{p}(t)}{dt^{2}}=\left[  eE_{\phi}^{E}(\mathbf{r}%
_{e}\left(  t\right)  ,t)\right]  .
\end{equation}
Thus for each charge, the proposed solution in Eq. (\ref{Mrphip}) indeed
satisfies the differential equation (\ref{Mr2dphi}), and also satisfies the
boundary conditions $\phi_{p}\left(  0\right)  =0$ and $\dot{\phi}_{p}\left(
0\right)  =0$ for the circular orbit of the charge. \ 

\subsection{Energy Gain from Random Radiation}

For each charge, the energy delivered by the quadrupole multipole of the
random radiation during a short time interval $\tau$ involving many
oscillations but little change in the mechanical angular momentum $J_{\phi}$
of each particle is
\begin{align}
W\left(  \tau\right)   &  =\int_{0}^{\tau}dt\left[  eE_{\phi}^{E}%
(\mathbf{r}_{e}\left(  t\right)  ,t)\right]  \left[  r_{e}\frac{d\phi_{p}%
(t)}{dt}\right]  \nonumber\\
&  =\int_{0}^{\tau}dt\left[  eE_{\phi}^{E}(\mathbf{r}_{e}\left(  t\right)
,t)\right]  \left\{  \frac{1}{M}\int_{0}^{t^{\prime}=t}dt^{\prime}\left[
eE_{\phi}^{E}(\mathbf{r}_{e}\left(  t^{\prime}\right)  ,t^{\prime})\right]
\right\}  \nonumber\\
&  =\frac{e^{2}}{M}\int_{0}^{\tau}dt\int_{0}^{t^{\prime}=t}dt^{\prime}\left[
E_{\phi}^{E}(\mathbf{r}_{e}\left(  t\right)  ,t)\right]  \left[  E_{\phi}%
^{E}(\mathbf{r}_{e}\left(  t^{\prime}\right)  ,t^{\prime})\right]
.\label{WPtauP}%
\end{align}
The integrand in Eq. (\ref{WPtauP}) is \textit{symmetric} under interchange of
$t$ and $t^{\prime}$. \ \ But then we can carry through Born's symmetrizing.
\ The double integral over the isosceles triangular region for $t$ and
$t^{\prime}$ in Eq. (\ref{WPtauP}) is the same as obtained by integrating
first in $t$ from $t^{\prime}$ to $\tau$, and then in $t^{\prime}$ from $0$ to
$\tau$ as%

\begin{align}
&  \int_{0}^{\tau}dt\int_{0}^{t^{\prime}=t}dt^{\prime}\left[  E_{\phi}%
^{E}(\mathbf{r}_{e}\left(  t\right)  ,t)\right]  \left[  E_{\phi}%
^{E}(\mathbf{r}_{e}\left(  t^{\prime}\right)  ,t^{\prime})\right] \nonumber\\
&  =\int_{0}^{\tau}dt^{\prime}\int_{t=t^{\prime}}^{t=\tau}dt\left[  E_{\phi
}^{E}(\mathbf{r}_{e}\left(  t\right)  ,t)\right]  \left[  E_{\phi}%
^{E}(\mathbf{r}_{e}\left(  t^{\prime}\right)  ,t^{\prime})\right]  .
\end{align}
However, we may interchange the prime and unprime labels and add half the
expressions to obtain for each charge%
\begin{equation}
W\left(  \tau\right)  =\frac{e^{2}}{2M}\int_{0}^{t=\tau}dt\int_{0}^{t^{\prime
}=\tau}dt^{\prime}\left[  E_{\phi}^{E}(\mathbf{r}_{e}\left(  t\right)
,t)\right]  \left[  E_{\phi}^{E}(\mathbf{r}_{e}\left(  t^{\prime}\right)
,t^{\prime})\right]  , \label{Wtaue2}%
\end{equation}
where now both time integrals are from $0$ to $\tau$. \ 

\subsection{Averaging Over the Random Phases of the Zero-Point Radiation}

Now we need the \textit{average} energy absorbed in the short time $\tau$
involving many passages over the charged-particle orbit. \ Thus, equation
(\ref{Wtaue2}) referring to the quadrupole multipole for each charge becomes%
\begin{equation}
\left\langle W\left(  \tau\right)  \right\rangle _{\theta^{E}}=\frac{e^{2}%
}{2M}\int_{0}^{\tau}dt\int_{0}^{\tau}dt^{\prime}\left\langle \left[  E_{\phi
}^{E}(\mathbf{r}_{e}\left(  t\right)  ,t)\right]  \left[  E_{\phi}%
^{E}(\mathbf{r}_{e}\left(  t^{\prime}\right)  ,t^{\prime})\right]
\right\rangle _{\theta^{E}}, \label{AvWtau}%
\end{equation}
where we average over the random phases $\theta^{E}$ of the driving radiation. \ 

The $\mathbf{r}_{e}\left(  t\right)  $ in the argument of the electric field
refers to the periodic motion of each charged particle in the Coulomb
potential and has nothing to do with the phases of the driving zero-point
radiation. \ However, we must maintain the $\mathbf{r}_{e}\left(  t\right)  $
dependence which corresponds to periodic particle motion in time. \ Then for
each charge, the $\phi$-component of the quadrupole multipole electric field
due to zero-point radiation can be rewritten as
\begin{align}
\left[  E_{\phi}^{E}(\mathbf{r}_{e}\left(  t\right)  ,t)\right]   &
=\operatorname{Re}E_{\phi}^{E}\left(  \mathbf{r}_{e}\left(  t\right)
,0\right)  \exp\left[  -i\omega^{E}t+i\theta^{E}\right] \nonumber\\
&  =E_{\phi}^{E}\left(  \mathbf{r}_{e}\left(  t\right)  ,0\right)  \cos\left[
\omega^{E}t-\theta^{E}\right]  .
\end{align}

Random radiation can be treated by introducing random phases for the waves.
\ However, there is already a phase in the problem, that of each charged
particle at opposite ends of a diameter in the circular orbit. \ Thus we must
be careful to preserve the randomness of the radiation waves relative to the
phase already present in the orbital motion. \ In equation (\ref{AvWtau}),
there are two phases corresponding to the two different radiation waves at
times $t$ and $t^{\prime}$ involved in the integrals. \ However, if the two
phases are different, the integrals over many periods will give vanishing
results. \ Thus the only contribution which survives the integration is that
where the quadrupole radiation field is matched with itself. \ However, this
radiation field will still retain its random phase compared with the phase of
the two orbiting charges. \ 

We are interested in averaging over the random phases $\theta_{n^{\prime
}l^{\prime}m^{\prime}}^{E}$ to obtain $\left\langle W\left(  \tau\right)
\right\rangle _{\theta^{E^{\prime}}}$ as in Eq. (\ref{AvWtau}). \ We require%
\begin{equation}
\left\langle \cos\left[  \theta_{nlm}^{E}\right]  \cos\left[  \theta
_{n^{\prime}l^{\prime}m^{\prime}}^{E}\right]  \right\rangle _{\theta
^{E^{\prime}}}=\left\langle \sin\left[  \theta_{nlm}^{E}\right]  \sin\left[
\theta_{n^{\prime}l^{\prime}m^{\prime}}^{E}\right]  \right\rangle
_{\theta^{E^{\prime}}}=\frac{1}{2}\delta_{nlm,n^{\prime}l^{\prime}m^{\prime}},
\end{equation}
and%
\begin{equation}
\left\langle \cos\left[  \theta_{nlm}^{E}\right]  \sin\left[  \theta
_{n^{\prime}l^{\prime}m^{\prime}}^{E}\right]  \right\rangle _{\theta
^{E^{\prime}}}=0.
\end{equation}

\ Then for the electric quadrupole multipole radiation, averaging and then
summing to eliminate the $\delta_{nlm,n^{\prime}l^{\prime}m^{\prime}}$, we
have%
\begin{align}
&  \left\langle \left[  E_{\phi}^{E}(\mathbf{r}_{e}\left(  t\right)
,t)\right]  \left[  E_{\phi}^{E}(\mathbf{r}_{e}\left(  t^{\prime}\right)
,t^{\prime})\right]  \right\rangle _{\theta^{E^{\prime}}}\nonumber\\
&  =\left\langle \sum\nolimits_{nlm}E_{\phi}^{E}\left(  \mathbf{r}_{e}\left(
t\right)  ,0\right)  \cos\left[  \omega^{E}t-\theta_{nlm}^{E}\right]
\sum\nolimits_{n^{\prime}l^{\prime}m^{\prime}}E_{\phi}^{E}\left(
\mathbf{r}_{e}\left(  t^{\prime}\right)  ,0\right)  \cos\left[  \omega
^{E}t^{\prime}-\theta_{n^{\prime}l^{\prime}m^{\prime}}^{E}\right]
\right\rangle _{\theta^{E^{\prime}}}\nonumber\\
&  =\sum\nolimits_{nlm}E_{nlm-\phi}^{E}(\mathbf{r}_{e}\left(  t\right)
,0)\sum\nolimits_{n^{\prime}l^{\prime}m^{\prime}}E_{n^{\prime}l^{\prime
}m^{\prime}-\phi}^{E}(\mathbf{r}_{e}\left(  t^{\prime}\right)  ,0)\cos\left[
\omega_{n^{\prime}l^{\prime}m^{\prime}}^{E}\left(  t-t^{\prime}\right)
\right]  \frac{1}{2}\delta_{nlm,n^{\prime}l^{\prime}m^{\prime}}\nonumber\\
&  =\frac{1}{2}\sum\nolimits_{nlm}\left[  E_{nlm-\phi}^{E}(\mathbf{r}%
_{e}\left(  t\right)  ,0)\right]  \left[  E_{nlm-\phi}^{E}(\mathbf{r}%
_{e}\left(  t^{\prime}\right)  ,0)\right]  \cos\left[  \omega_{nlm}^{E}\left(
t-t^{\prime}\right)  \right]  .
\end{align}
Then the average energy absorbed by each charge in Eq. (\ref{AvWtau}) is
\begin{align}
&  \left\langle W^{E}\left(  \tau\right)  \right\rangle _{\theta^{E^{\prime}}%
}\nonumber\\
&  =\frac{e^{2}}{2M}\int_{0}^{\tau}dt\int_{0}^{\tau}dt^{\prime}\left\langle
\left[  E_{\phi}^{E}(\mathbf{r}_{e}\left(  t\right)  ,t)\right]  \left[
E_{\phi}^{E}(\mathbf{r}_{e}\left(  t^{\prime}\right)  ,t^{\prime})\right]
\right\rangle \nonumber\\
&  =\frac{e^{2}}{2M}\int_{0}^{\tau}dt\int_{0}^{\tau}dt^{\prime}\frac{1}{2}%
\sum\nolimits_{nlm}\left[  E_{nlm-\phi}^{E}(\mathbf{r}_{e}\left(  t\right)
,0)\right]  \left[  E_{nlm-\phi}^{E}(\mathbf{r}_{e}\left(  t^{\prime}\right)
,0)\right]  \cos\left[  \omega_{nlm}^{E}\left(  t-t^{\prime}\right)  \right]
. \label{wtauq}%
\end{align}
Now we can use the $cos(a-b)$ expansion to rewrite Eq. (\ref{wtauq}) for each
charge as%
\begin{align}
\left\langle W^{E}\left(  \tau\right)  \right\rangle  &  =\frac{e^{2}}{4M}%
\sum\nolimits_{nlm}\left\{  \left[  \int_{0}^{\tau}dt\left[  E_{nlm-\phi}%
^{E}(\mathbf{r}_{e}\left(  t\right)  ,0)\right]  \cos\left[  \omega_{nlm}%
^{E}t\right]  \right]  ^{2}\right. \nonumber\\
&  +\left.  \left[  \int_{0}^{\tau}dt\left[  E_{nlm-\phi}^{E}(\mathbf{r}%
_{e}\left(  t\right)  ,0)\right]  \sin\left[  \omega_{nlm}^{E}t\right]
\right]  ^{2}\right\}  \label{avWtaue2}%
\end{align}

The two charged particles are going around with frequency $\omega_{e}$ in a
circular orbit of radius $r_{e}$, so that each has a speed is $\omega_{e}%
r_{e}=v_{e}$. \ Accordingly, the first integral in the electric quadrupole
multipole field in Eq. (\ref{avWtaue2}) requires the expansion involving the
position of each charge as%

\begin{equation}
E_{\phi}^{E}\left[  \mathbf{r}_{e}\left(  t\right)  ,0\right]  =\sum
\nolimits_{n^{\prime}=1}^{\infty}E_{n^{\prime}lm-\phi}^{E}\left(
k_{n^{\prime}lm}^{E}r_{e}\right)  \cos\left[  \omega_{e-n}t\right]
\end{equation}
where $l=2,m=\pm l$. \ Then including the time behavior of the electric field,
we have%
\begin{align}
&  \int_{0}^{\tau}dt\left[  E_{n^{\prime}lm-\phi}^{E}(\mathbf{r}_{e}\left(
t\right)  ,0)\right]  \cos\left[  \omega_{n^{\prime}lm}^{E}t\right]
\nonumber\\
&  =\int_{0}^{\tau}dt\left[  \sum\nolimits_{n^{\prime}lm}^{\infty}%
E_{n^{\prime}lm-\phi}^{E}\left(  k_{n^{\prime}lm}^{E}r\right)  \cos\left[
\omega_{e}t\right]  \right]  \cos\left[  \omega_{n^{\prime}lm}^{E}t\right]
\nonumber\\
&  =\int_{0}^{\tau}dt\sum\nolimits_{n^{\prime}lm}^{\infty}E_{n^{\prime}%
lm-\phi}^{E}\left(  k_{n^{\prime}lm}^{E}r\right)  \frac{1}{2}\left\{
\cos\left[  \left(  \omega_{e}-\omega_{nlm}^{E}\right)  t\right]  +\cos\left[
\left(  \omega_{e}+\omega_{nlm}^{E}\right)  t\right]  \right\} \nonumber\\
&  =\sum\nolimits_{n^{\prime}lm}^{\infty}E_{n^{\prime}lm-\phi}^{E}\left(
k_{n^{\prime}lm}^{E}r\right)  \frac{1}{2}\left\{  \frac{\sin\left[  \left(
\omega_{e}-\omega_{n^{\prime}lm}^{E}\right)  \tau\right]  }{\left(  \omega
_{e}-\omega_{n^{\prime}lm}^{E}\right)  }+\frac{\sin\left[  \left(  \omega
_{e}+\omega_{n^{\prime}lm}^{E}\right)  \tau\right]  }{\left(  \omega
_{e}+\omega_{n^{\prime}lm}^{E}\right)  }\right\}  ,
\end{align}
where $\omega_{e}$ is the frequency of each charged particle in its orbit and
the electric field of frequency $\omega_{nlm}^{E}$ is the quadrupole multipole
field. \ For a circular orbit, the expression $\left[  E_{nlm-\phi}^{E}%
(r_{e},\pi/2,0,0)\right]  $ does not involve time $t$ and so can be taken
outside the time integral. \ 

For a very large spherical enclosure of radius $\mathsf{R}$, the sum turns
into an integral, and we may use Eqs. (\ref{Ezprt}) and (\ref{aEnlm2}) giving
for the resonances $\omega_{e}\approxeq\omega_{n^{\prime}lm}^{E}$
\begin{align}
&  \left\langle W_{l,m}^{E}\left(  \tau\right)  \right\rangle \nonumber\\
&  =\frac{e^{2}}{4M}\sum\nolimits_{nlm}\left[  E_{nlm-\phi}^{E}(r_{e}%
,\pi/2,0,0)\right]  ^{2}\left\{  \left[  \int_{0}^{\tau}dt\cos\left[  \left(
\omega_{e}-\omega_{n^{\prime}l^{\prime}m^{\prime}}^{E}\right)  t\right]
\right]  ^{2}\right.  \nonumber\\
&  +\left.  \left[  \int_{0}^{\tau}dt\sin\left[  \left(  \omega_{e}%
-\omega_{n^{\prime}l^{\prime}m^{\prime}}^{E}\right)  t\right]  \right]
^{2}\right\}  \nonumber\\
&  =\frac{e^{2}}{4M}\left(  \int_{0}^{\infty}d\omega^{E}\frac{\mathsf{R}}{\pi
c}\right)  \left[  m^{2}\left[  \frac{1}{x}\left(  \frac{d\left[
xj_{l}\left(  x\right)  \right]  }{d\left(  x\right)  }\right)  \right]
_{x=mv/c}Y_{lm}\frac{a_{lm}^{E}}{\sqrt{l\left(  l+1\right)  }}\right]
_{\theta=\pi/2,\phi=0}^{2}\nonumber\\
&  \times\left\{  \left[  \int_{0}^{\tau}dt\cos\left[  \left(  \omega
_{e}-\omega^{E}\right)  t\right]  \right]  ^{2}+\left[  \int_{0}^{\tau}%
dt\sin\left[  \left(  \omega_{e}-\omega^{E}\right)  t\right]  \right]
^{2}\right\}  \nonumber\\
&  =\frac{e^{2}}{4M}\left(  \int_{0}^{\infty}d\omega^{E}\frac{\mathsf{R}}{\pi
c}\right)  m^{4}\left[  \frac{1}{x}\left(  \frac{d\left(  xj_{l}\left(
x\right)  \right)  }{dx}\right)  \right]  _{x=mv/c}^{2}\frac{\left\vert
Y_{lm}\right\vert ^{2}}{l\left(  l+1\right)  }\left[  \frac{16\pi\left(
\omega_{nl}^{E}\right)  ^{3}}{c^{2}\mathsf{R}}\left\langle J_{rad}\left(
\omega_{nl}^{E}\right)  \right\rangle \right]  \nonumber\\
&  \times\left\{  \left[  \int_{0}^{\tau}dt\cos\left[  \left(  \omega
_{e}-\omega^{E}\right)  t\right]  \right]  ^{2}+\left[  \int_{0}^{\tau}%
dt\sin\left[  \left(  \omega_{e}-\omega^{E}\right)  t\right]  \right]
^{2}\right\}  \nonumber\\
&  =\frac{4\pi e^{2}}{M}\left(  \int_{0}^{\infty}d\omega^{E}\frac{\mathsf{R}%
}{\pi c}\right)  m^{4}\left[  \frac{1}{x}\left(  \frac{d\left(  xj_{l}\left(
x\right)  \right)  }{dx}\right)  \right]  _{x=mv/c}^{2}\frac{\left\vert
Y_{lm}\right\vert ^{2}}{l\left(  l+1\right)  }\left[  \frac{\left(
\omega_{nl}^{E}\right)  ^{3}}{c^{2}\mathsf{R}}\left\langle J_{rad}\left(
\omega_{nl}^{E}\right)  \right\rangle \right]  \nonumber\\
&  \times\left\{  \left[  \frac{\sin^{2}\left[  \left(  \omega_{e}-\omega
^{E}\right)  \tau\right]  }{\left(  \omega_{e}-\omega^{E}\right)  ^{2}%
}\right]  +\left[  \frac{\left(  1-\cos\left[  \left(  \omega_{e}-\omega
^{E}\right)  \tau\right]  \right)  ^{2}}{\left(  \omega_{e}-\omega^{E}\right)
^{2}}\right]  \right\}  `
\end{align}

For a single spherical wave mode $l=2,m=\pm l$ at frequency $\omega^{E}$, we
integrate over $\omega^{E}$ as
\begin{equation}
\int_{0}^{\infty}d\omega^{E}\left[  \frac{\sin^{2}\left[  \left(  \omega
_{e}-\omega^{E}\right)  \tau\right]  }{\left(  \omega_{e}-\omega^{E}\right)
^{2}}\right]  \approxeq\int_{-\infty}^{\infty}d\omega^{E}\left[  \frac
{\sin^{2}\left[  \left(  \omega_{e}-\omega^{E}\right)  \tau\right]  }{\left(
\omega_{e}-\omega^{E}\right)  ^{2}}\right]  =\pi\tau,
\end{equation}
and%
\begin{equation}
\int_{0}^{\infty}d\omega^{E}\left[  \frac{\left(  1-\cos\left[  \left(
\omega_{e}-\omega^{E}\right)  \tau\right]  \right)  ^{2}}{\left(  \omega
_{e}-\omega^{E}\right)  ^{2}}\right]  \approxeq\int_{-\infty}^{\infty}%
d\omega^{E}\left[  \frac{\left(  1-\cos\left[  \left(  \omega_{e}-\omega
^{E}\right)  \tau\right]  \right)  ^{2}}{\left(  \omega_{e}-\omega^{E}\right)
^{2}}\right]  =\pi\tau.
\end{equation}

For the ground state where $\omega^{E}\approxeq\omega_{e}$, the integral in
$\omega^{E}$ collapses because of the resonance at $\omega^{E}=\omega_{e},$
and we have
\begin{equation}
P_{l,m}^{gain-E}=\frac{e^{2}}{M}\frac{\left(  \omega^{E}\right)  ^{3}}{c^{3}%
}\left\langle \left[  J_{rad}\left(  \omega_{nl}^{E}\right)  \right]
\right\rangle \left\{  8\pi\frac{m^{4}\left[  Y_{lm}\right]  ^{2}}{l\left(
l+1\right)  }\left[  \frac{1}{x}\frac{d\left[  xj_{l}(x)\right]  }{dx}\right]
_{x=mv/c}^{2}\right\}  .
\end{equation}
This expression is to be compared to the power lost in the same radiation mode
given by Burko\cite{Burko}%
\begin{align}
&  P_{l,m}^{loss-E}=8\pi\frac{e^{2}}{c^{3}}m^{4}\omega_{e}^{4}r_{e}^{2}%
\frac{l\left(  l+1\right)  }{\left(  2l+1\right)  ^{2}}\left[  Y_{l,m}\left(
\pi/2,0\right)  \right]  ^{2}\left[  \frac{1}{l+1}j_{l+1}\left(  m\omega
_{e}r_{e}/c\right)  -\frac{1}{l}j_{l-1}\left(  m\omega_{e}r_{e}/c\right)
\right]  ^{2}\nonumber\\
&  =M_{0}c^{2}\left(  \frac{M_{0}c^{3}}{e^{2}}\right)  \frac{\left[
e^{2}/\left(  J_{\phi}c\right)  \right]  ^{8}}{1-\left[  e^{2}/\left(
J_{\phi}c\right)  \right]  ^{2}}\left\{  8\pi\frac{m^{4}\left[  Y_{l,m}\left(
\pi/2,0\right)  \right]  ^{2}}{l\left(  l+1\right)  }\left[  \frac{1}{x}%
\frac{d\left[  xj_{l}(x)\right]  }{dx}\right]  _{x=mv/c}^{2}\right\}
\end{align}

The condition for energy balance for each of the charges is the requirement
that the power lost equals the power gained,
\begin{equation}
P_{l,m}^{loss-E}=P_{l,m}^{gain-E}.
\end{equation}
\ We remove the common factors of
\begin{equation}
8\pi\frac{m^{4}\left\vert Y_{lm}\right\vert ^{2}}{l\left(  l+1\right)
}\left[  \frac{1}{x}\frac{d\left[  xj_{l}(mx)\right]  }{dx}\right]
_{x=v/c}^{2},\label{Common}%
\end{equation}
and simplify using
\begin{equation}
\omega^{E}=\omega_{e}=\frac{Mc^{3}}{e^{2}}\left(  \frac{e^{2}}{\hbar
c}\right)  ^{3}=\frac{Me^{4}}{\hbar^{3}}%
\end{equation}
to find\cite{Goldstein}
\begin{equation}
J_{\phi}=J_{rad}=\hbar.\label{JeJrad}%
\end{equation}
In the ground state, the connection in Eq. (\ref{JeJrad}) gives energy balance
for each particle between the loss of energy due to emission of radiation as
quadrupole radiation and the gain of energy from the quadrupole multipole
driving by zero-point radiation.

\section{Estimate for Ground State of Helium}

We start with the helium ion He$^{+}$.\ Thus the helium nucleus has two plus
charges $q=2e$, and so the lowest nonrelativistic approximation for the energy
is obtained by taking an electron as orbiting in a circular hydrogen-like
orbit around the nucleus
\begin{align}
U_{H_{e}^{+}}  &  =-\frac{M_{0}\left(  Ze^{2}\right)  ^{2}}{2\left(
J_{2}\right)  ^{2}}=-\frac{M_{0}\left(  2e^{2}\right)  ^{2}}{2\left(
J_{2}\right)  ^{2}}=4\left(  -\frac{M_{0}\left(  e^{2}\right)  ^{2}}{2\left(
J_{2}\right)  ^{2}}\right) \nonumber\\
&  =4\left(  -13.6eV\right)  =-54.4eV. \label{UHepM}%
\end{align}
Now in the classical point of view, we want to add a second electron in a
fashion so as to minimized the electrostatic repulsion between the two
electrons. \ The simplest arrangement is to place the second electron on the
opposite side of the nucleus from the first electron. \ Then, ignoring the
repulsion between the two electrons, we have an equal contribution to the
energy from the second electron, so that%
\begin{equation}
U_{He-no}\approxeq2\left(  -54.4eV\right)  =-108.8eV,
\end{equation}
where the subscript \textquotedblleft$no$\textquotedblright\ refers to
\textquotedblleft no interaction between the two electrons.\textquotedblright

The next approximation for helium involves the approximate repulsion between
the two electrons. \ The two electrons are on opposite sides of the circular
orbit, and, in order to minimize the repulsion, they clearly must remain
opposite to each other as the orbital motion carries them around the nucleus.
\ Thus, the distance between the two electrons is $2r_{0He}$. \ However, the
radius of the helium-ion ground-state orbit follows from
\begin{equation}
M_{0}\frac{v^{2}}{r_{_{H_{e}^{+}}}}=\frac{Ze^{2}}{r_{_{H_{e}^{+}}}^{2}}\text{
\ \ or \ \ }v_{_{H_{e}^{+}}}=\frac{Ze^{2}}{J_{_{H_{e}^{+}}}}=\frac{2e^{2}%
}{J_{_{H_{e}^{+}}}},
\end{equation}
where $J_{H_{e}^{+}}=r_{H_{e}^{+}}Mv_{H_{e}^{+}}$. \ The orbital (angular)
frequency $\omega_{He}$ is given by%
\begin{equation}
\frac{\partial U_{H^{+}e}}{\partial J_{2}}=\omega_{H_{e}^{+}}=\frac
{M_{0}\left(  2e^{2}\right)  ^{2}}{J_{2}^{3}}=4\frac{M_{0}e^{4}}{\hbar^{3}%
}=4\omega_{Bohr},\label{wHep}%
\end{equation}
giving a radius
\begin{equation}
r_{_{H_{e}^{+}}}=\frac{v_{_{H_{e}^{+}}}}{\omega_{_{H_{e}^{+}}}}=\frac{2e^{2}%
}{J_{_{H_{e}^{+}}}}\left(  \frac{J_{2}^{3}}{M_{0}\left(  2e^{2}\right)  ^{2}%
}\right)  =\frac{J_{2}^{2}}{M_{0}2e^{2}}=\frac{1}{2}\frac{\hbar^{2}}%
{M_{0}e^{2}}=\frac{1}{2}r_{Bohr}.\label{rHep}%
\end{equation}
Thus we find that the diameter of the orbital motion is%
\begin{equation}
2r_{_{H_{e}^{+}}}=\frac{J_{2}^{2}}{M_{0}e^{2}}=\frac{\hbar^{2}}{M_{0}e^{2}}.
\end{equation}
The separation of the two electrons is just the Bohr radius. \ Accordingly,
the classical electrostatic repulsion energy between the two electrons of the
same sign is twice the attractive energy of the Bohr hydrogen atom%
\begin{equation}
PE_{cl}=\frac{e^{2}}{2r_{_{H_{e}^{+}}}}=\frac{M_{0}e^{4}}{\hbar^{2}%
}=2\left\vert -13.6eV\right\vert =27.2eV.
\end{equation}

Accordingly, the approximate energy of the ground state of helium within
classical electrodynamics with classical zero-point radiation is%
\begin{equation}
U_{He-cl}\approxeq-108.8eV+27.2eV=-81.6eV.\text{ \ \ classical}\label{UHeCl}%
\end{equation}
This classical electromagnetic approximation given in Eq. (\ref{UHeCl}) seems
satisfactory since it differs from the accepted $U_{He}=-79.0$ ground-state
value for helium by only $2.6/79=3\%.$ \ 

On the other hand, Griffiths' quantum text\cite{Griffiths} poses the
calculation of this approximation as a moderately hard problem. \ The quantum
solution for the electrostatic repulsion between the electrons gives $PE_{q}=$
$34.0eV$ with an associated approximated energy for the ground state of helium%
\[
U_{He-q}\approxeq-108.8eV+34.0eV=-74.8eV.\text{ \ \ quantum}%
\]
Thus the difference of the modern quantum result from the accepted value of
$-79.0eV$ is $4.2eV,$ giving an error of $4.2/79=5\%$. \ In this instance, the
classical electrodynamic calculation is distinctly easier and slightly closer
to the accepted value.

\section{Conclusions}

It appears as though the inclusion of Lorentz-invariant classical
electromagnetic zero-point radiation will give an average energy balance for
the ground state of the helium atom within classical electromagnetic theory.
\ The lowest approximation for the classical result for the average energy of
the helium atom is close to but not the same as that given by modern quantum
theory, and differs by a few percent from the accepted experimental value. \ 

Thus the ground state for helium does not seem outside the purview of
classical electrodynamics including classical electromagnetic zero-point
radiation. \ The helium ground state can be added to the successful results of
classical theory which gives results depending on Planck's constant $\hbar$:
Casimir forces, van der Waals forces, oscillator specific heats, blackbody
radiation, superfluid behavior, and the states of the hydrogen
atom.\cite{Classical}

July 2, 2026 \ \ \ \ \ \ \ \ \ \ \ Helium4.tex
\end{document}